\documentclass{article}

\usepackage{arxiv}

\usepackage[utf8]{inputenc}
\usepackage[T1]{fontenc}
\usepackage{hyperref}
\usepackage{url}
\usepackage{booktabs}
\usepackage{amsfonts}
\usepackage{amsmath}
\usepackage{microtype}
\usepackage{cleveref}
\usepackage{graphicx}
\usepackage[numbers,sort&compress]{natbib}
\usepackage{doi}
\usepackage{float}
\usepackage[table]{xcolor}

\title{Architecture is All You Need: Improving LLM Recommenders by Dropping the Text}

\author{
  Kevin Foley \\
  Comcast Applied AI \\
  \texttt{kevin\_foley@comcast.com}
  \and
  Shaghayegh Agah \\
  Comcast Applied AI \\
  \texttt{shaghayegh\_agah@comcast.com}
  \and
  \\
  Kavya Priyanka Kakinada \\
  Comcast Applied AI \\
  \texttt{kavyapriyanka\_kakinada@comcast.com}
}

\renewcommand{\shorttitle}{Architecture is All You Need}

\hypersetup{
pdftitle={Architecture is All You Need: Improving LLM Recommenders by Dropping the Text},
pdfsubject={Recommender Systems, LLM, Sequential Recommendation},
pdfauthor={Kevin Foley, Shaghayegh Agah, Kavya Priyanka Kakinada},
pdfkeywords={Multitask, Recommendation, Personalization, Large Language Models, Content Retrieval},
}

\begin{document}
\maketitle

\begin{abstract}
In recent years, there has been an explosion of interest in the applications of large pre-trained language models (PLMs) to recommender systems, with many studies showing strong performance of PLMs on common benchmark datasets. PLM-based recommender models benefit from flexible and customizable prompting, an unlimited vocabulary of recommendable items, and general ``world knowledge'' acquired through pre-training on massive text corpora. While PLM-based recommenders show promise in settings where data is limited, they are hard to implement in practice due to their large size and computational cost. Additionally, fine-tuning PLMs to improve performance on collaborative signals may degrade the model's capacity for world knowledge and generalizability. We propose a recommender model that uses the architecture of large language models (LLMs) while reducing layer count and dimensions and replacing the text-based subword tokenization of a typical LLM with discrete tokens that uniquely represent individual content items. We find that this simplified approach substantially outperforms both traditional sequential recommender models and PLM-based recommender models at a tiny fraction of the size and computational complexity of PLM-based models. Our results suggest that the principal benefit of LLMs in recommender systems is their architecture, rather than the world knowledge acquired during extensive pre-training.

\end{abstract}


\section{Introduction}
Recent research in the field of recommender systems has begun to explore the capabilities of large language models (LLMs) as bases for fine-tuned next-item recommendation models. These LLM-based recommenders have shown strong performance and techniques like Supervised Fine-Tuning (SFT)~\cite{bao2025bi} and Direct Preference Optimization (DPO)~\cite{rafailov2023direct, chen2024softmax} have been used to further tune language models' outputs to produce high-quality recommendations. The LLM model architectures used in large pretrained language models (PLMs) offer many advantages over earlier generations of sequential model architectures. Above all, LLMs are excellent pattern learners. Decoder-only transformer architectures with causal attention masks excel at next-token prediction in settings with long context windows and complex dependencies and interactions between input tokens. This is an advantage in language modeling that can just as easily be applied to the next-item prediction task that is at the center of most recommendation systems.

\begin{figure}[t]
  \centering
  \includegraphics[width=0.28\textwidth]{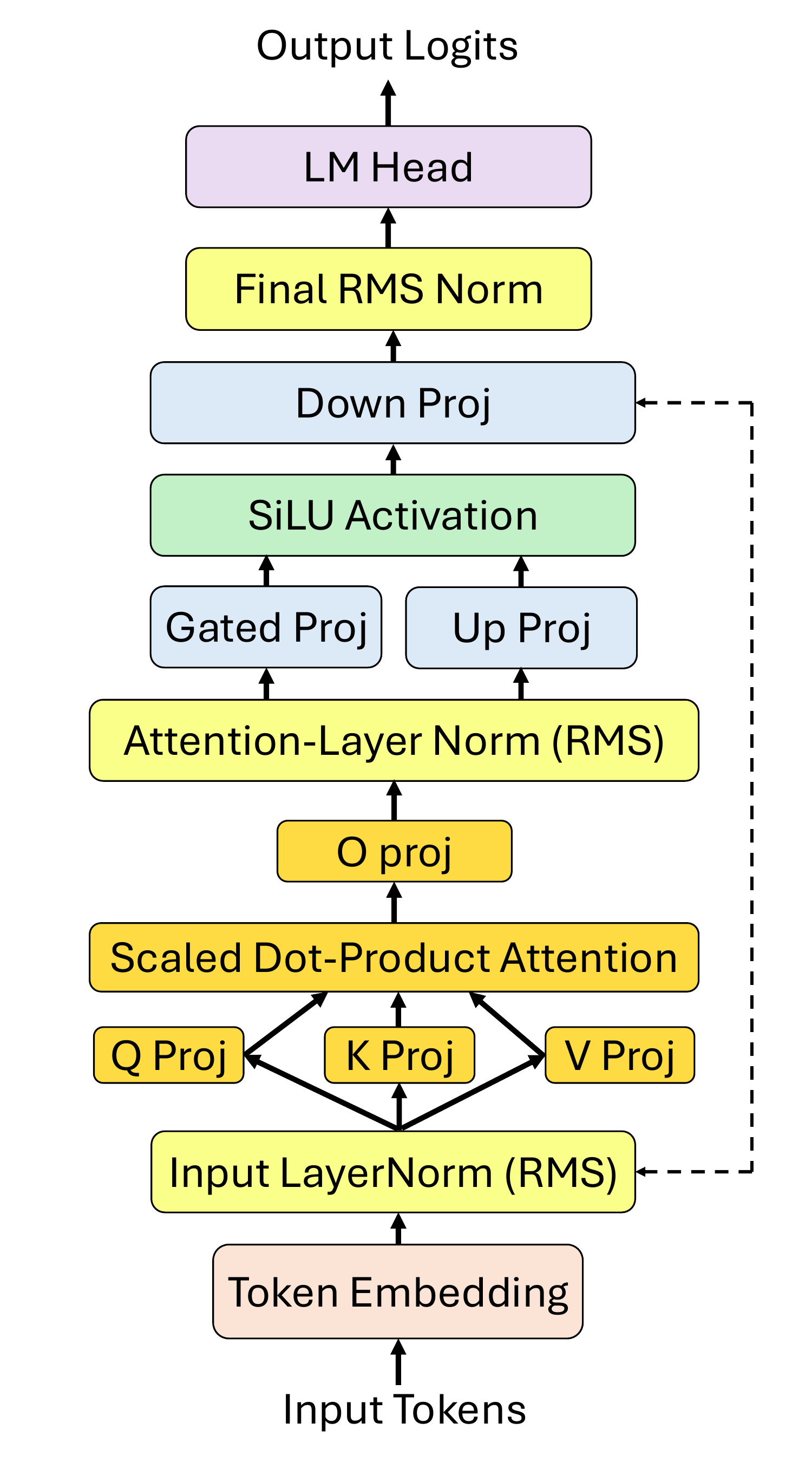}
  \caption{Model architecture. With the exception of the SiLU activation to reduce parameter count, the architecture is the same as that of the Llama3 family of models.}
  \label{fig:model_arch}
\end{figure}

In addition to architectural advantages of LLMs, PLMs can contribute to recommender systems with \textit{world knowledge}. They excel at generalizing and synthesizing information from massive collections of text, and can be quickly adapted to new tasks through fine tuning or in-context learning and few-shot prompting \cite{brown2020fewshot}. When used in recommendation systems, PLMs have shown a promising ability to generate high quality recommendations on little or no training data and to recommend cold start items not previously seen in the training data. Several recent PLM recommender models -- BigRec~\cite{bao2025bi}, TALLRec~\cite{bao2023tallrec}, SPRec~\cite{gao2025sprec}, S-DPO~\cite{chen2024softmax}, RoseDPO~\cite{liao2024rosepo}, LLara~\cite{liao2024llara}, and GPT4Rec~\cite{li2023gpt4rec} -- have shown strong performance against popular benchmarks. On the other hand, PLM recommender systems are computationally expensive, often thousands of times larger than sophisticated transformer-based recommender models. They require many more tokens to be processed per recommended item in both input prompts and generated output. The world knowledge of PLMs has been shown to degrade after finetuning \cite{luo2025catastrophicforgetting}, meaning that attempts to improve recommendation performance through fine tuning could limit the advantage of the PLM's world knowledge. Additionally, a recent study \cite{dipalma2025llms} finds that the PLMs used for recommender models may have been exposed during training to common benchmark datasets, calling into question the strong results that have been reported by these models.

In this work, we develop LSRec, a lightweight sequential recommender, and compare the benefits of the LLM architecture to the benefits of PLM's world knowledge. LSRec shares the same causal decoder-only transformer architecture as the Llama family of open-weight PLMs without using text input or pre-trained weights. We compare our model performance with SASRec, a leading transformer-based next-item recommender operating directly on item tokens, and with BigRec, a recent fine-tuned PLM recommender that shows strong performance on limited training data.

\section{Methodology}

Traditional next-item recommendation models typically take as input a sequence of item ids that a user has interacted with, and either score a single candidate item or output a vector of scores in the style of a multilabel classifier. PLM-based models typically take the title, year, and possibly additional metadata for each item watched, format the sequence of watched titles and metadata into a string as part of a larger text prompt explaining the recommendation task, and generate the title and year of an output item. The generated text serves as the model’s recommendation and must be mapped to item ids.

We propose a hybrid approach combining the architecture of a causal, decoder-only transformer model with the item-level tokenization of a traditional sequential recommendation model. We base our model on the Llama 3 architecture \cite{grattafiori2024llama}. This is preferred over alternative LLM architectures due to its relatively simple design and wide adoption in applied research papers, including in PLM recommendation model research.

We reduce model size by specifying a shorter context window, fewer layers and hidden dimensions, a single key-value attention head instead of the multiple key-value heads in a typical LLM, and limiting vocabulary size to the full set of ids in the item vocabulary plus a handful of special tokens for multi-task training (explained in the following section). Other configuration differences between a pretrained Llama model and LSRec are shown in Table~\ref{tab:model_comparison}. To further reduce the size of our models, we use sigmoid linear unit (SiLU) activation instead of the Swish-gated Linear Unit (SwiGLU) activation function used in the Llama family of LLMs. We specify three models ranging in size from 768 thousands to 7.1 million parameters. With these modifications, LSRec ranges from around 150 to 1,000 times smaller than even the smallest pretrained Llama models.
Simplified model architecture is shown in Figure~\ref{fig:model_arch}.

\begin{table}[t!]
\centering
\small
\hspace*{-0.25in}
\begin{tabular}{lcccc}
\toprule
\textbf{Parameter} & \textbf{Llama 3.2-1b} & \textbf{LSRec-768k} & \textbf{LSRec-2.2M} & \textbf{LSRec-7.1M} \\
\midrule
params & 1B & 768k & 2.2M & 7.1M \\
attn dropout & 0.0 & 0.2 & 0.2 & 0.2 \\
hidden dims & 2048 & 64 & 128 & 384 \\
intermediate dims & 8192 & 128 & 384 & 512 \\
context length & 131k & 200 & 200 & 200 \\
attn heads & 32 & 1 & 2 & 2 \\
layers & 16 & 2 & 4 & 4 \\
kv heads & 8 & 1 & 1 & 1 \\
vocab size & 128k & 10k & 10k & 10k \\
\bottomrule
\end{tabular}
\vspace{0.1in}
\caption{Key Configuration Differences vs. Llama3.2-1b}
\label{tab:model_comparison}
\end{table}

\subsection{Data preparation}

We focus our analysis on recommendations in the movie domain and train and evaluate various LSRec versions on the MovieLens-10M dataset, a widely-used benchmark dataset in recommender system papers. The dataset consists of 10 million user ratings from 72,000 users across 10,000 movies, with genre tags for each individual movie.
User-item interactions are ordered by timestamp into user-item sequences. Users with fewer than five interacted items are dropped from the dataset, and sequences are truncated to a maximum length of 200 items. We prepare our training data for supervised fine tuning (SFT) using a simple prompt template that includes a handful of special tokens.

\subsection{Multi-task training}

Previous studies have shown that training deep learning models on additional tasks in correlated domains can improve performance on the primary task~\cite{mehrotra2020bandit}. With that in mind, beside the movie recommend task, we create additional related tasks to better leverage the flexibility of our promptable transformer model as well as the additional data available in the MovieLens dataset. We define the following five tasks:

\textbf{movie:} Recommend a movie to a user based on past history of user-item interactions.

\textbf{genre:} Recommend a film genre to a user based on past history of user-item interactions.

\textbf{rating:} Predict a user's star rating for a given item based on past history of user-item interactions.

\textbf{movie by genre:} Recommend a movie to a user based on a sequence of past user-item interactions and a genre in the prompt.

\textbf{movie by rating:} Recommend a movie to a user based on a sequence of past user-item interactions and a star rating included in the prompt.

The context signal for every task consists only of that user's past history of item interactions. Tasks are specified by the special TASK token followed by a task id token. Target data (the token to be predicted) and additional arguments are included at the end of the prompt after special tokens ARGUMENTS and START. The prompt format is as follows:

\small
$$\text{BOS }\mathbf{m_i, ..., m_i}\text{ TASK }\mathbf{task\_id}\text{ ARGUMENTS } \mathbf{arg\_id}\text{ START }\mathbf{m_{target}}\text{ EOS}$$
\normalsize

\noindent where \textbf{m\textsubscript{i}, ..., m\textsubscript{i}} are the sequence of past user-item interactions, \textbf{task\_id} is a special task token and ARGUMENTS \textbf{arg\_id} optionally specifies additional arguments such as a genre or rating prompt. Note that because the model is not pre-trained on a text corpus, the choice of text for token labels is arbitrary. The detail prompt for each task is provided in Table~\ref{fig:prompt_examples}.

\noindent\textbf{Example:} Consider a user with a viewing history consisting of movies: (1234, The Matrix), (5678, Inception), (9012, Blade Runner 2049), (3456, Ex Machina), and (7890, Interstellar). The model receives the id sequence \begin{verbatim}BOS M1234 M5678 M9012 M3456 M7890 TASK RECOMMEND START \end{verbatim} as input, where each id corresponds to its respective movie title in the dataset mapping. LSRec is trained to predict an appropriate next movie id recommendation, such as M2468 (Arrival), demonstrating its ability to capture latent preference patterns for science fiction films through the learned embeddings of movie identifiers.

\begin{table}[h]
\centering
\small
\begin{tabular}{p{1.2cm}p{13cm}}
\toprule
\textbf{Task} & \textbf{Prompt Format} \\
\midrule
movie & BOS \{m\_1, ..., m\_n\} TASK RECOMMEND START \{target\} EOS \\
genre & BOS \{m\_1, ..., m\_n\} TASK RECOMMEND\_GENRE START \{target\} EOS \\
rating & BOS \{m\_1, ..., m\_n\} TASK RATE\_MOVIE ARGUMENTS \{movie\} START \{target\} EOS \\
by genre & BOS \{m\_1, ..., m\_n\} TASK MOVIE\_BY\_GENRE ARGUMENTS \{genre\} START \{target\} EOS \\
by rating & BOS \{m\_1, ..., m\_n\} TASK RECOMMEND\_RATING ARGUMENTS \{rating\} START \{target\} EOS \\
\bottomrule
\end{tabular}
\vspace{0.1in}
\caption{LSRec Prompt Format for Multiple Tasks}
\label{fig:prompt_examples}

\end{table}

\section{Experimental Setup}

To evaluate the performance of the recommendation system, we generated next-item recommendations for all users in the MovieLens dataset and evaluated recommendation quality based on the position of the ground truth target in the full set of recommended labels, excluding items already present in the user-item interaction history. We follow existing research by reserving the most recently interacted item in the user-item interaction sequence as the test dataset target, the second-most recent item as the evaluation target, and the third-most-recent item as the training target.

\subsection{Metrics}

We use three metrics common in the literature on recommender systems: Hit Rate (HR), Normalized Discounted Cumulative Gain (NDCG) and Diversity Ratio (Div). Div quantifies the uniqueness of recommendations across users. It is calculated as the number of unique items recommended to all test users divided by the total number of items in the vocabulary.

Although LSRec models have a generative architecture, ranked lists of recommended items were generated by taking the top K logits from a single generated token in the target position (and not by generating a sequence of K tokens). While the model is also capable of sequential generation, we leave the evaluation of sequential generations for future work.

\begin{table*}[!h]
    \centering
    \footnotesize
    \setlength{\tabcolsep}{3pt}
    \begin{tabular}{llcccccccccc}
    \toprule
   \textbf{} & \textbf{Model} & \textbf{Params} & \textbf{HR@1} &  \textbf{Div@1} & \textbf{NDCG@10} & \textbf{HR@10} & \textbf{Div@10} & \textbf{NDCG@20} & \textbf{HR@20} & \textbf{Div@20}\\
    \midrule
\textbf{Replicated}   & SASRec & 3.3M & 0.0142 & 0.1722 & 0.0607 & 0.1288  & 0.4656 & 0.0821 & 0.2140 & 0.5810\\
  \textbf{Baselines}  & BigRec-1024 & 7B & 0.0098 & 0.1241 & 0.0147  &  0.0214 & 0.3736 & 0.0171 & 0.0308 & 0.4959 \\
 &   BigRec-full & 7B & 0.0323 & 0.1708 & 0.0381  &  0.0456 & 0.4792 & 0.0400 & 0.0531 & 0.6146\\
 \midrule
\textbf{LSRec } &   LSRec-small & 768.3k &0.0360 & 0.1564 &  0.0991 & 0.1850 & 0.2813& 0.1206 & 0.2702 & 0.3331\\
\textbf{Versions} &  LSRec-small-mt & 768.3k &0.0584& 0.3016 & 0.1357 & 0.2383 & 0.5407& 0.1592 & 0.3317 & 0.6256\\
 &  LSRec-medium & 2.2M & 0.0741 & 0.3051 & 0.1616 & 0.2748 & 0.5118& 0.1861 & 0.3717 & 0.5801\\
 & LSRec-medium-mt & 2.2M &\textbf{0.0952}& 0.3744 & \textbf{0.1909} & \textbf{0.3129} & 0.6725& \textbf{0.2147}& \textbf{0.4075} & 0.7600\\
&   LSRec-large & 7.1M &0.0658& 0.4006 &0.1488 & 0.2578 & 0.6987 & 0.1725 & 0.3518 & 0.7798\\
&   LSRec-large-mt & 7.1M &0.0859& \textbf{0.4233} & 0.1772 & 0.2935 & \textbf{0.7133}& 0.2012 & 0.3887 & \textbf{0.7909}\\
    \bottomrule
    \end{tabular}
    \caption{Performance Comparison: Inference on all users and items. The best results are highlighted in bold.}
    \label{tab:main_results}

\end{table*}

\subsection{Baselines}
To compare our results, we replicated two baselines — SASRec~\cite{kang2018self}, an earlier generation sequential transformer-based model, and BigRec~\cite{bao2025bi}, a fine-tuned LLM model for the recommendation task. For other alignment techniques on LLMs, we are in the process of replication. Meanwhile, we compared our results to those reported in the SPRec~\cite{gao2025sprec} paper.

\textbf{SASRec:} We compared our model's performance against SASRec~\cite{kang2018self}, a sequential recommendation model based on self-attention mechanisms. SASRec learns user preferences from sequences of past interactions and is a strong baseline for capturing user preference patterns.

\textbf{BigRec:} We evaluate LSRec variations against BigRec~\cite{bao2025bi}, a PLM-based recommender fine-tuned to generate the title and year of the next predicted item and then retrieve the closest matches to that recommendation via KNN embeddings lookup. We chose BigRec as our PLM baseline because of its simplicity, strong performance on limited data and because the KNN lookup method is straightforward to replicate and easily generates a ranked list of recommended items grounded in the item vocabulary. We replicate the results reported in Bao \textit{et al.} study~\cite{bao2025bi} over just 1024 examples (BigRec-1024) and, for better comparison with our training scheme over the full dataset (BigRec-full).

\subsection{Training Details}

All models were trained using supervised fine-tuning (SFT) for up to 100 epochs with an early stopping patience of 20 for non-PLM based models and a patience of 5 for PLM-based models. Sequence packing was used to maximize training efficiency. Models were trained and evaluated on a single NVIDIA L40S GPU. All models and baselines were trained on the same MovieLens data. All evaluation metrics were measured using the output of our model and baselines to assess relative performance.

\section{Results and Discussion}
Training results are reported in Table~\ref{tab:main_results}. LSRec outperforms both the sequential transformer baseline (SASRec) and the PLM recommender baseline (BigRec) across all metrics with both standard and multi-task training while using far fewer parameters. The smallest LSRec model achieves an 80\% improvement in HR@1 over the PLM baseline with only 0.08\% of the PLM's parameter count, and a 311\% improvement over the transformer baseline at just 23\% of SASRec's parameter count. The mid-sized model performs best, achieving an 81\% improvement over the PLM baseline with 0.22\% of the parameters, and a 570\% improvement over the transformer baseline with 67\% of the parameters. Notably, HR and NDCG do not improve when increasing model size from 2.2 million-parameter medium model to the 7.1 million parameter large model. This means that a model with around 2 million parameters is sufficient for recommendation system settings with a dataset size similar to MovieLens-10M, and that larger models of this class are only beneficial when trained on datasets much larger than the current dataset.

Interestingly, LSRec achieves superior performance relative to SASRec. Both models share similar transformer architectures. The key differences between our model's Llama architecture and SASRec are: (1) Rotary Position Embeddings (LLama) vs. positional embeddings (SASRec), (2) Grouped query attention (LLama) vs. standard multi-head attention (SASRec), (3) Llama's relatively complex feedforward network design and and activation functions, (4) Different layer normalization, and finally, (5) SASRec is trained via negative sampling while our models were trained using supervised fine tuning. Additional investigation is needed to fully understand the factors driving these performance gains.

LSRec models also show a marked improvement over the PLM recommender baselines. Several factors may explain this performance difference. First, token-based recommenders use data more efficiently and are able to use much longer context windows for a given computational budget. In this study, SASRec and LSRec had context windows of up to 200 items, while the longer per-item token sequences and higher computational cost of PLM-based recommenders means that (following the procedure of the BigRec paper) input sequences were limited to ten items. A second advantage of token-based sequential recommenders over PLM-based recommenders is output reliability: PLM recommenders can hallucinate responses that do not exist in the vocabulary of available programs, and generating additional recommendations beyond the first item requires either a vector search (BigRec) or continued generation.

 \begin{table}[h!]
\centering
\small
\begin{tabular}{llcccc}
\toprule
 \textbf{} &\textbf{Model}   &\textbf{NDCG@5} & \textbf{HR@5} & \textbf{Div@5} \\
\midrule
\textbf{Replicated}  &SASRec              &0.0491 &   0.0844    & 0.0444 \\
\textbf{Baselines} &BIGRec-1024      & 0.0048 &   0.0060    & 0.0580 \\
                    &BIGRec-full       &  0.0305 &   0.0330 & 0.1022 \\
                    \midrule
\textbf{Baselines}   &SDPO            & 0.0258 &   0.0310   & 0.1816 \\
\textbf{in SPRec}       &SPRec           & 0.0319 &  0.0388    & \textbf{0.2806} \\
                    \midrule
\textbf{LSRec}  &  LSRec-small          & 0.0805 &   0.114   & 0.1054 \\
\textbf{Versions}    &  LSRec-small-mt        & 0.1094 &  0.152   & 0.1469 \\
                     &  LSRec-medium          & 0.1274 &   0.187   & 0.1493 \\
                     &  LSRec-medium-mt       & \textbf{0.1664} &  \textbf{0.231}    & 0.1591 \\
                     &   LSRec-large          & 0.1181 &   0.170   & 0.1686 \\
                     &  LSRec-large-mt        & 0.1409 &  0.196    & 0.1663 \\
\bottomrule
\end{tabular}
\begin{minipage}{3.8in}
\vspace{0.1in}
\caption{Performance Comparison: Inference on 1,000 randomly sampled users and all items. The best results are highlighted in bold.}
\label{tab:divratio_k5_comparison}
\end{minipage}

\end{table}

As shown in Table~\ref{tab:main_results}, the multi-task (mt) version of each model consistently outperforms the corresponding single-task variant. This performance gain is observed across all key evaluation metrics, indicating that jointly optimizing for multiple objectives allows the models to learn more robust and generalizable representations of items and user preferences. This finding aligns with prior research, where correlated multi-objective and multi-task approaches, when properly implemented, have been shown to outperform single-objective baselines~\cite{mehrotra2020bandit, chu2025multi}.

To compare LSRec’s performance against the DPO-tuned PLM recommender, we replicate the evaluation strategy introduced by Gao \textit{et al}~\cite{gao2025sprec}, and compute metrics for the top-5 recommendations over a random sample of 1,000 users. These results are compared with our replicated baseline and the two DPO baselines reported in the SPRec paper~\cite{gao2025sprec} (which we did not independently replicate), as shown in Table~\ref{tab:divratio_k5_comparison}. Based on our observations, LSRec outperforms the DPO-based recommenders in NDCG and HR, while the latter achieves better diversity.\footnote{The difference between diversity metrics in tables ~\ref{tab:main_results} and ~\ref{tab:divratio_k5_comparison} are due to the smaller sample size in ~\ref{tab:divratio_k5_comparison}, which reduces the number of recommended items.}

\section{Conclusion and Future Work}

A key lesson from this study is that while LLM \textit{architectures} offer powerful advantages over other transformer-based recommender models, the use of large pre-trained language models is not necessary to achieve these benefits, and in fact the performance of token-based LLM recommenders appears to be far superior to the performance of much larger PLM-based models in any data-rich setting. This is a convenient finding, because token-based LLM recommenders are much smaller, faster and less expensive than their PLM counterparts. Because our models use a Llama architecture, they are compatible with a wide range of post-training, optimization, and model serving frameworks available as part of the growing open-source ecosystem supporting open-weight LLM training and inference.

On the other hand, we observe that while our models outperform both the traditional and PLM baselines in terms of diversity ratio of recommended items, they fall short of the performance of DPO-tuned PLMs on this metric. We leave for future work the question of whether very small token-based LLM recommenders can be successfully optimized using standard LLM post-training techniques such as DPO.

Finally, while we have shown that multi-task training on related domains can considerably improve model performance, evaluating the performance of those additional tasks was beyond the scope of this paper. Future work in this area will include evaluations of these additional tasks against suitable benchmarks and exploring other applications for promptable token-based LLM recommenders beyond the next-item recommendation task.

\bibliographystyle{unsrtnat}
\bibliography{references}

\end{document}